\documentstyle[12pt,epsf]{article}
\begin{document}
\newcommand {\ee}{\end{equation}}
\newcommand {\bea}{\begin{eqnarray}}
\newcommand {\eea}{\end{eqnarray}}
\newcommand {\nn}{\nonumber \\}
\newcommand {\Tr}{{\rm Tr\,}}
\newcommand {\tr}{{\rm tr\,}}
\newcommand {\e}{{\rm e}}
\newcommand {\etal}{{\it et al.}}
\newcommand {\m}{\mu}
\newcommand {\n}{\nu}
\newcommand {\pl}{\partial}
\newcommand {\p} {\phi}
\newcommand {\vp}{\varphi}
\newcommand {\vpc}{\varphi_c}
\newcommand {\al}{\alpha}
\newcommand {\be}{\beta}
\newcommand {\ga}{\gamma}
\newcommand {\Ga}{\Gamma}
\newcommand {\x}{\xi}
\newcommand {\ka}{\kappa}
\newcommand {\la}{\lambda}
\newcommand {\La}{\Lambda}
\newcommand {\si}{\sigma}
\newcommand {\th}{\theta}
\newcommand {\Th}{\Theta}
\newcommand {\om}{\omega}
\newcommand {\Om}{\Omega}
\newcommand {\ep}{\epsilon}
\newcommand {\vep}{\varepsilon}
\newcommand {\na}{\nabla}
\newcommand {\del}  {\delta}
\newcommand {\Del}  {\Delta}
\newcommand {\mn}{{\mu\nu}}
\newcommand {\ls}   {{\lambda\sigma}}
\newcommand {\ab}   {{\alpha\beta}}
\newcommand {\half}{ {\frac{1}{2}} }
\newcommand {\third}{ {\frac{1}{3}} }
\newcommand {\fourth} {\frac{1}{4} }
\newcommand {\sixth} {\frac{1}{6} }
\newcommand {\sqg} {\sqrt{g}}
\newcommand {\fg}  {\sqrt[4]{g}}
\newcommand {\invfg}  {\frac{1}{\sqrt[4]{g}}}
\newcommand {\sqZ} {\sqrt{Z}}
\newcommand {\gbar}{\bar{g}}
\newcommand {\sqk} {\sqrt{\kappa}}
\newcommand {\sqt} {\sqrt{t}}
\newcommand {\reg} {\frac{1}{\epsilon}}
\newcommand {\fpisq} {(4\pi)^2}
\newcommand {\Lcal}{{\cal L}}
\newcommand {\Ocal}{{\cal O}}
\newcommand {\Dcal}{{\cal D}}
\newcommand {\Ncal}{{\cal N}}
\newcommand {\Mcal}{{\cal M}}
\newcommand {\scal}{{\cal s}}
\newcommand {\Dvec}{{\hat D}}   
\newcommand {\dvec}{{\vec d}}
\newcommand {\Evec}{{\vec E}}
\newcommand {\Hvec}{{\vec H}}
\newcommand {\Vvec}{{\vec V}}
\newcommand {\Btil}{{\tilde B}}
\newcommand {\ctil}{{\tilde c}}
\newcommand {\Ftil}{{\tilde F}}
\newcommand {\Stil}{{\tilde S}}
\newcommand {\Ztil}{{\tilde Z}}
\newcommand {\altil}{{\tilde \alpha}}
\newcommand {\betil}{{\tilde \beta}}
\newcommand {\latil}{{\tilde \lambda}}
\newcommand {\ptil}{{\tilde \phi}}
\newcommand {\Ptil}{{\tilde \Phi}}
\newcommand {\natil} {{\tilde \nabla}}
\newcommand {\ttil} {{\tilde t}}
\newcommand {\Rhat}{{\hat R}}
\newcommand {\Shat}{{\hat S}}
\newcommand {\shat}{{\hat s}}
\newcommand {\Dhat}{{\hat D}}   
\newcommand {\Vhat}{{\hat V}}   
\newcommand {\xhat}{{\hat x}}
\newcommand {\Zhat}{{\hat Z}}
\newcommand {\Gahat}{{\hat \Gamma}}
\newcommand {\nah} {{\hat \nabla}}
\newcommand {\gh}  {{\hat g}}
\newcommand {\labar}{{\bar \lambda}}
\newcommand {\cbar}{{\bar c}}
\newcommand {\bbar}{{\bar b}}
\newcommand {\Bbar}{{\bar B}}
\newcommand {\psibar}{{\bar \psi}}
\newcommand {\chibar}{{\bar \chi}}
\newcommand {\bbartil}{{\tilde {\bar b}}}
\newcommand  {\vz}{{v_0}}
\newcommand {\intfx} {{\int d^4x}}
\newcommand {\inttx} {{\int d^2x}}
\newcommand {\change} {\leftrightarrow}
\newcommand {\ra} {\rightarrow}
\newcommand {\larrow} {\leftarrow}
\newcommand {\ul}   {\underline}
\newcommand {\pr}   {{\quad .}}
\newcommand {\com}  {{\quad ,}}
\newcommand {\q}    {\quad}
\newcommand {\qq}   {\quad\quad}
\newcommand {\qqq}   {\quad\quad\quad}
\newcommand {\qqqq}   {\quad\quad\quad\quad}
\newcommand {\qqqqq}   {\quad\quad\quad\quad\quad}
\newcommand {\qqqqqq}   {\quad\quad\quad\quad\quad\quad}
\newcommand {\qqqqqqq}   {\quad\quad\quad\quad\quad\quad\quad}
\newcommand {\lb}    {\linebreak}
\newcommand {\nl}    {\newline}

\newcommand {\vs}[1]  { \vspace*{#1 cm} }

\newcommand {\MPL}  {Mod.Phys.Lett.}
\newcommand {\NP}   {Nucl.Phys.}
\newcommand {\PL}   {Phys.Lett.}
\newcommand {\PR}   {Phys.Rev.}
\newcommand {\PRL}   {Phys.Rev.Lett.}
\newcommand {\CMP}  {Commun.Math.Phys.}
\newcommand {\JMP}  {Jour.Math.Phys.}
\newcommand {\AP}   {Ann.of Phys.}
\newcommand {\PTP}  {Prog.Theor.Phys.}
\newcommand {\NC}   {Nuovo Cim.}
\newcommand {\CQG}  {Class.Quantum.Grav.}


\font\smallr=cmr5
\def\ocirc#1{#1^{^{{\hbox{\smallr\llap{o}}}}}}
\def\ogamma{\ocirc{\gamma}{}}
\def\oM{{\buildrel {\hbox{\smallr{o}}} \over M}}
\def\osigma{\ocirc{\sigma}{}}

\def\overleftrightarrow#1{\vbox{\ialign{##\crcr
 $\leftrightarrow$\crcr\noalign{\kern-1pt\nointerlineskip}
 $\hfil\displaystyle{#1}\hfil$\crcr}}}
\def\overnab{{\overleftrightarrow\nabslash}}

\def\va{{a}}
\def\vb{{b}}
\def\vc{{c}}
\def\tilpsi{{\tilde\psi}}
\def\tbpsi{{\tilde{\bar\psi}}}

\def\Dslash{{}\hbox{\hskip2pt\vtop
 {\baselineskip23pt\hbox{}\vskip-24pt\hbox{/}}
 \hskip-11.5pt $D$}}
\def\nabslash{{}\hbox{\hskip2pt\vtop
 {\baselineskip23pt\hbox{}\vskip-24pt\hbox{/}}
 \hskip-11.5pt $\nabla$}}
\def\xislash{{}\hbox{\hskip2pt\vtop
 {\baselineskip23pt\hbox{}\vskip-24pt\hbox{/}}
 \hskip-11.5pt $\xi$}}
\def\leftnabla{{\overleftarrow\nabla}}

\def\delL{{\delta_{LL}}}
\def\delG{{\delta_{G}}}
\def\delc{{\delta_{cov}}}

\newcommand {\sqxx}  {\sqrt {x^2+1}}   
\newcommand {\gago}  {\gamma_5}
\newcommand {\Ktil}  {{\tilde K}}
\newcommand {\Ltil}  {{\tilde L}}
\newcommand {\Qtil}  {{\tilde Q}}
\newcommand {\Rtil}  {{\tilde R}}
\newcommand {\Kbar}  {{\bar K}}
\newcommand {\Lbar}  {{\bar L}}
\newcommand {\Qbar}  {{\bar Q}}
\newcommand {\Pp}  {P_+}
\newcommand {\Pm}  {P_-}
\newcommand {\GfMp}  {G^{5M}_+}
\newcommand {\GfMpm}  {G^{5M'}_-}
\newcommand {\GfMm}  {G^{5M}_-}
\newcommand {\Omp}  {\Omega_+}    
\newcommand {\Omm}  {\Omega_-}
\def\Aslash{{}\hbox{\hskip2pt\vtop
 {\baselineskip23pt\hbox{}\vskip-24pt\hbox{/}}
 \hskip-11.5pt $A$}}
\def\Rslash{{}\hbox{\hskip2pt\vtop
 {\baselineskip23pt\hbox{}\vskip-24pt\hbox{/}}
 \hskip-11.5pt $R$}}
\def\kslash{
{}\hbox       {\hskip2pt\vtop
                   {\baselineskip23pt\hbox{}\vskip-24pt\hbox{/}}
               \hskip-8.5pt $k$}
           }    
\def\qslash{
{}\hbox       {\hskip2pt\vtop
                   {\baselineskip23pt\hbox{}\vskip-24pt\hbox{/}}
               \hskip-8.5pt $q$}
           }    
\def\dslash{
{}\hbox       {\hskip2pt\vtop
                   {\baselineskip23pt\hbox{}\vskip-24pt\hbox{/}}
               \hskip-8.5pt $\partial$}
           }    
\def\dbslash{{}\hbox{\hskip2pt\vtop
 {\baselineskip23pt\hbox{}\vskip-24pt\hbox{$\backslash$}}
 \hskip-11.5pt $\partial$}}
\def\Kbslash{{}\hbox{\hskip2pt\vtop
 {\baselineskip23pt\hbox{}\vskip-24pt\hbox{$\backslash$}}
 \hskip-11.5pt $K$}}
\def\Ktilbslash{{}\hbox{\hskip2pt\vtop
 {\baselineskip23pt\hbox{}\vskip-24pt\hbox{$\backslash$}}
 \hskip-11.5pt ${\tilde K}$}}
\def\Ltilbslash{{}\hbox{\hskip2pt\vtop
 {\baselineskip23pt\hbox{}\vskip-24pt\hbox{$\backslash$}}
 \hskip-11.5pt ${\tilde L}$}}
\def\Qtilbslash{{}\hbox{\hskip2pt\vtop
 {\baselineskip23pt\hbox{}\vskip-24pt\hbox{$\backslash$}}
 \hskip-11.5pt ${\tilde Q}$}}
\def\Rtilbslash{{}\hbox{\hskip2pt\vtop
 {\baselineskip23pt\hbox{}\vskip-24pt\hbox{$\backslash$}}
 \hskip-11.5pt ${\tilde R}$}}
\def\Kbarbslash{{}\hbox{\hskip2pt\vtop
 {\baselineskip23pt\hbox{}\vskip-24pt\hbox{$\backslash$}}
 \hskip-11.5pt ${\bar K}$}}
\def\Lbarbslash{{}\hbox{\hskip2pt\vtop
 {\baselineskip23pt\hbox{}\vskip-24pt\hbox{$\backslash$}}
 \hskip-11.5pt ${\bar L}$}}
\def\Rbarbslash{{}\hbox{\hskip2pt\vtop
 {\baselineskip23pt\hbox{}\vskip-24pt\hbox{$\backslash$}}
 \hskip-11.5pt ${\bar R}$}}
\def\Qbarbslash{{}\hbox{\hskip2pt\vtop
 {\baselineskip23pt\hbox{}\vskip-24pt\hbox{$\backslash$}}
 \hskip-11.5pt ${\bar Q}$}}
\def\Acalbslash{{}\hbox{\hskip2pt\vtop
 {\baselineskip23pt\hbox{}\vskip-24pt\hbox{$\backslash$}}
 \hskip-11.5pt ${\cal A}$}}

\begin{flushright}
July 2001\\
hep-th/0107254 \\
US-01-03
\end{flushright}

\vspace{0.5cm}

\begin{center}
{\Large\bf 
Some Properties of Domain Wall Solution\\
in the Randall-Sundrum Model}

\vspace{1.5cm}
{\large Shoichi ICHINOSE
          \footnote{
E-mail address:\ ichinose@u-shizuoka-ken.ac.jp
                  }
}
\vspace{1cm}

{\large 
Laboratory of Physics, \\
School of Food and Nutritional Sciences, \\
University of Shizuoka,
Yada 52-1, Shizuoka 422-8526, Japan          
}

\end{center}
\vfill

{\large Abstract}\nl
Properties of the domain wall (kink) solution in the
5 dimensional Randall-Sundrum model 
are examined both {\it analytically} and
{\it numerically}. 
The configuration is derived by the bulk Higgs mechanism. 
We focus on 1) the convergence property of 
the solution, 
2) the stableness of the solution, 3) the non-singular
property of the Riemann curvature, 
4) the behaviours of the warp factor and the
Higgs field. It is found that 
the bulk curvature changes the sign around the surface
of the wall. 
We also present some {\it exact} solutions for 
two simple cases: 
a) the no potential case, b) the cosmological
term dominated case. Both solutions have the (naked) curvature singularity. 
We can regard the domain wall solution as a singularity
resolution of the exact solutions.

\vspace{0.5cm}

PACS NO:\ 
11.27.+d 
04.50.+h 
11.10.Kk 
11.25.Mj 
12.10.-g 
04.25.-g 
\nl
\section{Introduction}
There exist two standpoints
in the treatment of the domain world physics.
One is that the geometry should be {\it singular} and 
the domain is regarded as a {\it defect}. 
In the original work by Randall and Sundrum\cite{RS9905},
the walls stand on the fixed points of the $S_1/Z_2$ orbifold
in the form of $\del$-function. 
Recently
the renormalization of a bulk-boundary system
is discussed in this standpoint\cite{GW0104}
where the conical singularity at the center of the extra 2 dim
space is regarded as a domain of the vortex type.
The other standpoint is that the geometry should be {\it non-singular}. 
The domain is regarded as a {\it soliton}. 
This approach looks natural if the domain world is
"derived" from the more fundamental theory of the D-brane.
Similar situations occurred in the past literature, such as
the relation between
the Dirac string\cite{DirPR48} and 'tHooft-Polyakov string
\cite{tHNP74, PolJE74}.
(For the situation about the (topological) defect and the soliton
in relation to the domain world, see a good review \cite{CS97PRep}.) 
At present both standpoints look important to understand
the brane world physics. 
We take here the latter standpoint. 

In ref.\cite{SI0003}, the domain wall configuration
of the RS-model is realized as a soliton (kink) solution
in the bulk (5D) Higgs potential. It has some advantageous
points, compared with the $\del$-function description,
such as non-singularity and stability. 
The solution is obtained
in the form of the {\it infinite} power-series of 
some hyperbolic function. 
The convergence of the
coefficient-series is crucial 
for the boundary condition (b.c.) to be satisfied. 
It was checked by explicitly calculating the coefficients
at the 2nd order. 
We present here the 6th order calculation result, 
and reconfirm the convergence property further strongly. 
The main purpose of this paper is to strengthen the content
of ref.\cite{SI0003} by presenting the various results in
a concrete way. 
Some physical quantities, such as the bulk scalar curvature, 
are obtained. Very interestingly, the curvature changes its sign
near the "surface" of the domain wall.
In order to clarify the structure of
the solution, we first present some {\it exact} solutions for
simple cases. They clearly show the origin of some
integration constants and free parameters. 
These exact solutions have (naked) singularities. 
They tell us
the Higgs potential is important for the non-singular property
of the configuration. It plays the role of singular resolution.

We consider 
one-wall model which was considered in \cite{RS9906}.
An interesting {\it stable} (kink) solution
exists for a {\it family} of vacua. As explained
in \cite{SI0003}, 
the solution does not miss the key points
of the original one. Similar analysis was successfully done
in the 6 dim model\cite{SI0103}.

As some recent related works, we find \cite{ST0103, BCY0105}.

In Sec.2, the RS domain wall model is explained. The exact
solution is presented for the no-potential case in Sec.3,
and for the cosmological-constant dominated case in Sec.4.
The general case of the Higgs potential
is examined in Sec.5, where the domain-wall
solution is obtained as a one-parameter family of 
kink solutions. 
It is the {\it analytical} solution. 
The concrete values of parameters and coefficients 
are obtained in the 6th order calculation. 
{\it Numerical} analysis also
confirms the obtained solution.

\section{Randall-Sundrum model with the bulk Higgs field}
We consider, as the brane world, 
the following 5D gravity-Higgs theory.
\begin{eqnarray}
S[G_{AB},\Phi]=\int d^5X\sqrt{-G} (-\half M^3\Rhat
-\half G^{AB}\pl_A\Phi\pl_B\Phi-V(\Phi))\com\nn
V(\Phi)=\frac{\la}{4}(\Phi^2-{v_0}^2)^2+\La\com
\label{model1}
\end{eqnarray}
where $X^A (A=0,1,2,3,4)$ is the 5D coordinates and we also use
the notation $(X^A)\equiv (x^\m,y), \m=0,1,2,3.$
$X^4=y$ is the extra axis which is taken to be a space coordinate.
The signature of the 5D metric $G_{AB}$ is $(-++++)$. 
$\Phi$ is a 5D Higgs (scalar) field, $G=\det G_{AB}$, $\Rhat$ is the
5D Riemannian scalar curvature. $M$ and $V(\Phi)$ are the 5D Planck mass 
and the Higgs potential respectively. 
The three parameters $\la,\vz$ and $\La$ in $V(\Phi)$ are called here
{\it vacuum parameters}. 
$\la(>0)$ is a coupling, $v_0(>0)$ 
is the Higgs field vacuum expectation value, and $\La$ is the 5D cosmological constant. 

Assuming the Poincar{\' e} invariance in the brane,
the line element can be written as
\begin{eqnarray}
{ds}^2=\e^{-2\si(y)}\eta_\mn dx^\m dx^\n+{dy}^2
\equiv G_{AB}dX^AdX^B\com
\label{model2}
\end{eqnarray}
where $\eta_\mn=\mbox{diag}(-1,1,1,1)$. $(G_{AB})$ is explicitly
written as
\begin{eqnarray}
(G_{AB})=\left(  \begin{array}{cc}
                 \e^{-2\si}\eta_\mn,& 0\\
                 0,&   1             
                 \end{array}
         \right)\com\q
         \sqrt{-G}=\e^{-4\si}\pr
\label{model3}
\end{eqnarray}
The 5D Einstein equation gives us
\begin{eqnarray}
-6M^3(\si')^2=-\half (\Phi')^2+V\com\label{sol2a}\\
3M^3\si''=(\Phi')^2\com\label{sol2b}
\end{eqnarray}
where "$'$=$d/dy$" and we consider the case that $\Phi$ depends only 
on the extra coordinate $y$;\ $\Phi=\Phi(y)$. 
The above equations are 
1) translation invariant ($y\ra y+$ const.);\ 
2) Z$_2$ symmetric 
($y\ra -y, \si'\ra -\si', \Phi\ra\pm\Phi $);\ 
3)even with respect to the $\Phi$-reflection
($\Phi\change -\Phi$). 
Besides they are 
4) global scale invariant,
when some vacuum parameters change appropriately;
\begin{eqnarray}
y\ra ky\com\q
\la\ra\frac{\la}{k^2}\com\q
\La\ra\frac{\La}{k^2}\com\q
\vz\ra\vz\com
\label{model4}
\end{eqnarray}
where $k$ is a constant. 
This invariance says the scale of $y$ can be adjusted
by the scaling of $\la$ and $\La$. Note that
the scaling power is independent of their mass-dimensions. 
\footnote{
The mass-dimensions of the vacuum parameters, 
$\la$, $\La$ and $\vz$, are
(mass)$^{-1}$, (mass)$^{5}$, and (mass)$^{3/2}$
respectively. 
}

Eq.(\ref{sol2b})
gives an important positivity relation,
\begin{eqnarray}
\si''=\frac{1}{3M^3}(\Phi')^2\geq 0\com\nn
3M^3\{ \si'|_{y=y_2}-\si'|_{y=y_1}\}=
\int^{y_2}_{y_1}(\Phi')^2 dy\geq 0\com\q
y_1<y_2
\com\label{sol3b}
\end{eqnarray}
where non-singularity of $\si''$ is assumed in the region
$y_1<y<y_2$. This relation will serve as a consistency check
of the solutions.\footnote
{For a general argument about the consistency of the domain
world configuration, see \cite{GKL0011}.
}
We will also focus on the (non)singularity of the bulk curvature: 
$\Rhat=-8\si''+20{\si'}^2$.

As the extra space (the fifth dimension), we take 
the real number space ${\bf R}=(-\infty,+\infty)$.
This is a simplified version of the original RS-model\cite{RS9905} 
and was examined in the subsequent work\cite{RS9906}.

\section{Exact Solution for the No Potential Case}
Let us consider the case of no potential, $V=0$;\ 
$\la=\La=0$. 
Then eq.(\ref{sol2a},\ref{sol2b}) reduce to
\begin{eqnarray}
-6M^3(\si')^2=-\half (\Phi')^2\com\q
3M^3\si''=(\Phi')^2\pr\label{nopo2}
\end{eqnarray}
$\si(y)$ is solved as
\begin{eqnarray}
\si'=\frac{1}{A-4y}\com\q
\si=-\fourth\ln \frac{|A-4y|}{B}\com\q B>0\com
\label{nopo3}
\end{eqnarray}
where $A$($-\infty<A<\infty$) and $B$($>0$) are integration
constants. 
The constant, $A$, comes from the {\it translation invariance}
of (\ref{nopo2}). 
$B$ comes from
the {\it global scale invariance} of 
(\ref{nopo2}). 
The line element is given by
\begin{eqnarray}
{ds}^2=\sqrt{\frac{|A-4y|}{B}}\eta_\mn dx^\m dx^\n+{dy}^2
\pr
\label{nopo4}
\end{eqnarray}
$\Phi'$ and $\Phi$ are solved as
\begin{eqnarray}
\Phi'=\pm\frac{\sqrt{12M^3}}{A-4y}\com\q
\Phi=\mp\frac{\sqrt{3M^3}}{2}\ln \frac{|A-4y|}{C}\com\q C>0\com
\label{nopo5}
\end{eqnarray}
where $C$($>0$) is another integration constant.
The plural signs come from the evenness of (\ref{nopo2}) under
the "$\Phi$-reflection":\ $\Phi\change -\Phi$. 
The Higgs field, $\Phi(y)$, does {\it not} 
go to a constant in the asymptotic region $|y|\ra\infty$, 
which should be compared with other solutions obtained later.

The 5D Riemann scalar curvature is obtained as
\begin{eqnarray}
\Rhat=
\frac{-12}{(A-4y)^2}<0
\com
\label{nopo6}
\end{eqnarray}
which is 1) negative definite, 2) {\it singular} at $y=A/4$ and
3) vanishes for $|y|\ra \infty$. 
The metric (\ref{nopo4}) has no horizon, hence this curvature
singularity is a {\it naked} one.

The obtained solution has unwanted properties and 
can not be used as the brane world model.
This model is too simple. 
It has, however, 
some common or comparative features to the more realistic
solutions of later sections in some points such as
1) the appearance of some integration constants
in relation to some symmetries of the field equations, 
2) the plural signs, 
3) (naked) curvature singularity, 4) no horizon.
Furthermore a suggestive relation to the R-S metric can be
found by considering the region :\ 
$|A|\gg |y|$. In this case the metric (\ref{nopo4}) reduces to
\begin{eqnarray}
{ds}^2\approx\sqrt{\frac{|A|}{B}}\e^{-\frac{2}{A}y}
\eta_\mn dx^\m dx^\n+{dy}^2\com\q
|A|\gg |y|
\com
\label{nopo7}
\end{eqnarray}
which looks like the RS-type metric although we cannot take
$|y|\ra \infty$.

\section{Exact Solution for the Cosmological Term Dominated Case}
Let us examine a little more general case,that is, the cosmological
term dominated case:\ $\la=0$. 
\begin{eqnarray}
-6M^3(\si')^2=-\half (\Phi')^2+\La\com\label{cos1a}\\
3M^3\si''=(\Phi')^2\pr\label{cos1b}
\end{eqnarray}
We have interest in the case: $\Phi\ra\mbox{const.}$ as
$y\ra\pm\infty$.
Then, from (\ref{cos1a}), $\La\leq 0$. 

From above equations, we get
$3M^3\si''/(12M^3{\si'}^2+2\La)=1$, which can be integrated as
\begin{eqnarray}
\si'=\om\frac{1+A\e^{8\om y}}{1-A\e^{8\om y}}\com\q
\om\equiv \sqrt{\frac{-\La}{6M^3}}
\com\label{cos2}
\end{eqnarray}
where $A$ is an integration constant ($-\infty<A<\infty$)
which comes from the
{\it translation invariance} of (\ref{cos1a},\ref{cos1b}). \nl
\nl
(i) $A<0$\nl
The solution (\ref{cos2}) can be written as
\begin{eqnarray}
\si'=-\om\tanh\ 4\om(y-y_*)\com\q |A|\equiv \e^{-8\om y_*}
\pr\label{cos3}
\end{eqnarray}
It is attractive that this solution is non-singular.  
But it
contradicts with the positivity relation (\ref{sol3b}).
Hence we conclude, in this case (i), there does {\it not} exist 
a consistent solution.\nl
\nl
(ii) $A=0$\nl
This case is solved as
\begin{eqnarray}
\si'=\om\com\q
\si=\om y+B\com\q
\Phi=\vz
\com\label{cos4}
\end{eqnarray}
where $B$ and $\vz$ are another integration constants.
The 5D curvature is a {\it positive constant} everywhere.
\begin{eqnarray}
\Rhat =\frac{10}{3}\frac{(-\La)}{M^3}>0
\pr\label{cos5}
\end{eqnarray}
The geometry is Anti de Sitter space.\nl
\nl
(iii) $A>0$\nl
(\ref{cos2}) is singular at $y=-(\ln A)/8\om \equiv y_*$. The
solutions are obtained as
\begin{eqnarray}
\si'=-\om\coth\ 4\om(y-y_*)\com\nn
\Phi'=\pm\frac{\sqrt{-2\La}}{\sinh (4\om(y-y_*))}\com\nn
\Phi=\pm\frac{\sqrt{3M^3}}{2}\ln|\tanh (2\om(y-y_*))|+\vz
\com\label{cos6}
\end{eqnarray}
where $\vz$ is an integration constant. 
Fields asymptotically behave as
$\si'\ra \mp\om,\ \Phi\ra\vz$ when $y-y_*\ra\pm\infty$. 
The 5D scalar curvature is
obtained as
\begin{eqnarray}
\Rhat=\frac{-\La}{M^3}\{
\frac{10}{3}-\frac{2}{ \{\sinh(4\om(y-y_*))\}^2 }
                   \}
\com\label{cos7}
\end{eqnarray}
which is singular at $y=y_*$. 
It asymptotically behaves as
$\Rhat\ra -\frac{10}{3}\frac{\La}{M^3}$ 
when $|y-y_*|\ra\infty$. 
Integrating $\si'$ in eq.(\ref{cos6}), the line element is obtained as
\begin{eqnarray}
{ds}^2=B\sqrt{|\sinh(4\om(y-y_*))|}\,\eta_\mn dx^\m dx^\n+dy^2
\com\label{cos8}
\end{eqnarray}
where $B$ is another integration constant. 
From this result, we see there is
no horizon. The curvature singularity at $y=y_*$ is the 
{\it naked} one.
In the region far from the singularity ($4\om|y-y_*|\gg 1$), the line
element can be approximated as
${ds}^2\approx (B/\sqrt{2})
\e^{2\om|y-y_*|}\eta_\mn dx^\m dx^\n+dy^2$, which is
similar to the RS metric except the sign of the exponent. 

We can check the solution of (iii) goes to (ii) as
$A\ra +0$($y_*\ra\infty$). 
We can further check the solution (iii) is continuously connected to
the no potential case of Sec.3 by taking $\La\ra -0$. 
The parameter $A$ in Sec.3 ($\equiv A^{\mbox{no-pot}}$) is related
to $A$ of (\ref{cos2}) as
$1/A^{\mbox{no-pot}}=\om (1+A)/(1-A)$. 
In the point that $\Phi, \si'$ and $\Rhat$ become
(non-zero) constants in the asymptotic region, 
the solution of this section approaches,
compared with Sec.3, a realistic one. 

In Sec.3 and 4, we have obtained the {\it exact} solutions.
They become RS-type solutions for special regions of $y$.
( It suggests that further "deformation" of the potential 
makes us find the domain wall solution. Indeed it will do so. )
The solutions, however, still have a bad property of 
(naked) curvature singularity. To seek a non-singular solution, 
we must take into account all vacuum parameters
$\la,\vz$ and $\La$. That is the following subject.
\section{Domain Wall Solution}
Let us solve the 5D Einstein equations (\ref{sol2a},\ref{sol2b})
for the general case of vacuum parameters.
We impose the following asymptotic behaviour 
(boundary condition)
for the Higgs field $\Phi(y)$.
\begin{eqnarray}
\Phi(y)\ra\pm v_0\com\q
y\ra\pm\infty
\label{sol3}
\end{eqnarray}
This means $\Phi'\ra 0$, and from (\ref{sol2b}), $\si''\ra 0$.
From this result and (\ref{sol3b}), we are led to
$\si'\ra\pm\om, \si\ra\om |y|$ as $y\ra\pm\infty$, 
where $\om(>0)$ is some
constant. 
It can be fixed, by considering $y\ra\pm\infty$
in (\ref{sol2a}), as
\begin{eqnarray}
\om=\sqrt{\frac{-\La}{6}}M^{-\frac{3}{2}}\com\q
\La\leq 0\com
\label{sol4}
\end{eqnarray}
where we see the sign of $\La$ must be {\it negative}, that is,
the 5D geometry must be {\it anti de Sitter} in the
asymptotic regions.
We may set $M=1$ without ambiguity. (Only when it is necessary,
we explicitly write down $M$-dependence.)

We notice, in the results of previous sections, 
$\si'$ and $\Phi$ behave in a comparative way. 
Let us take the following form for $\si'(y)$ and $\Phi(y)$
as a solution.
\begin{eqnarray}
\si'(y)=\om\sum_{n=0}^\infty\frac{c_{2n+1}}{(2n+1)!}\{\tanh (ky+l)\}^{2n+1}\com\nn
\Phi(y)=v_0\sum_{n=0}^\infty\frac{d_{2n+1}}{(2n+1)!}\{\tanh (ky+l)\}^{2n+1}\com
\label{sol5}
\end{eqnarray}
where $c$'s and $d$'s are coefficient-constants to be determined. 
\footnote{
Normalization of $c_{2n+1}$ is different from that of ref.\cite{SI0003}.
The relation is $c_{2n+1}=(k/\om)\times c_{2n+1}$ of 
ref.\cite{SI0003}.
}
The odd-power terms are only taken here using the $Z_2$ symmetry
of (\ref{sol2a},\ref{sol2b}) and the boundary conditions
explained above. 
The free parameter $l$ comes from 
the {\it translation invariance} of (\ref{sol2a}) and (\ref{sol2b}). 
A {\it new mass scale} $k(>0)$ is introduced here 
as a free parameter to make 
the quantity  $k y$ dimensionless. The freedom
comes from the global scale invariance of (\ref{sol2a}) and (\ref{sol2b}).
The physical meaning of $1/k$ 
is the {\it thickness} of the domain wall. 
The parameter $k$ plays a central role
in the dimensional reduction scenario\cite{SI0003}.
The distortion of 5D space-time geometry by the
existence of the domain wall should be small so that 
the {\it quantum} effect of 5D gravity
can be ignored and the present {\it classical} analysis is valid. This requires
the condition\cite{RS9905}
\begin{eqnarray}
k\ll M\pr
\label{sol6}
\end{eqnarray}
Besides $M=1$, we can also take $k=1$ without ambiguity
(keeping the relation (\ref{sol6}) in mind).
The coefficient-constants
$c$'s and $d$'s have the following constraints, 
\begin{eqnarray}
1=\sum_{n=0}^\infty\frac{c_{2n+1}}{(2n+1)!}
\com\q
1=\sum_{n=0}^\infty\frac{d_{2n+1}}{(2n+1)!}\com
\label{sol7}
\end{eqnarray}
which are obtained by considering the asymptotic behaviours
$y\ra \pm\infty$ in (\ref{sol5}). 

We first obtain the recursion relations between
the expansion coefficients,
from the field equations (\ref{sol2a}) and (\ref{sol2b}).
For $n\geq 2$, they are given by \cite{SI0003}
\begin{eqnarray}
\frac{c_{2n+1}}{(2n)!}-\frac{c_{2n-1}}{(2n-2)!}=
\frac{\vz^2}{\sqrt{-3\La/2}}
({D'}_n-2{D'}_{n-1}+{D'}_{n-2})\com\nn
C_{n-1}=-\frac{\vz^2}{2\La}({D'}_n-2{D'}_{n-1}+{D'}_{n-2})
+\frac{\la}{4}\frac{\vz^4}{\La}(E_{n-2}-2D_{n-1})\ ,
\label{sol8}
\end{eqnarray}
where 
\begin{eqnarray}
D_n=\sum_{m=0}^{n}\frac{d_{2n-2m+1}d_{2m+1}}{(2n-2m+1)!\,(2m+1)!}\com\q
{D'}_n=\sum_{m=0}^{n}\frac{d_{2n-2m+1}d_{2m+1}}{(2n-2m)!\,(2m)!}\ ,\nn
C_n=\sum_{m=0}^{n}\frac{c_{2n-2m+1}c_{2m+1}}{(2n-2m+1)!\,(2m+1)!}\com\q
E_n=\sum_{m=0}^{n}D_{n-m}D_m\pr
\label{sol8b}
\end{eqnarray}
The first few terms, $(c_1,d_1),(c_3,d_3)$, 
are explicitly given as
\begin{eqnarray}
d_1=\pm\frac{\sqrt{2}}{\vz}\sqrt{\La +\frac{\la \vz^4}{4}}\com\q
c_1=\frac{2}{3\sqrt{-\La/6}}(\La+\frac{\la \vz^4}{4})\com\nn
\frac{d_3}{d_1}=2+
\{ \frac{8}{3}(\La +\frac{\la \vz^4}{4})-\la\vz^2\}
\ ,\ 
\frac{c_3}{c_1}=2+
\{ \frac{16}{3}(\La +\frac{\la \vz^4}{4})-2\la\vz^2\}
\ ,
\label{sol9}
\end{eqnarray}
where $\pm$ sign in $d_1$ reflects $\Phi\change -\Phi$ symmetry
in (\ref{sol2a}) and (\ref{sol2b}).
We take the positive one in the following.
We can confirm that the above relations, (\ref{sol8}) and (\ref{sol9}), 
determine all $c$'s and $d$'s
recursively in the order of increasing $n$. 
In (\ref{sol8}) and (\ref{sol9}), $M=k=1$ is taken for
simplicity. Their dependence is easily recovered by
$\La\ra\La /k^2M^3,\ \la\ra\la M^3/k^2,\ \vz\ra\vz/\sqrt{M^3}$.
Note that 
all coefficients, derived above, are solved and are described by
the three (dimensionless) vacuum parameters. 
Among the 3 parameters, there exist 2 constraints from (\ref{sol7}).
Hence the present solution is {\it one-parameter family} solution.
 
In order for this solution to make sense, as seen from the expression
for $d_1$, the 5D cosmological term $\La$ should be bounded both 
from {\it below} and from {\it above}.
\begin{eqnarray}
-\frac{\la\vz^4}{4}<\La<0\pr
\label{sol10}
\end{eqnarray}
The presence of this relation says the non-singular solution
presented here cannot continuously connect with the singular
solutions of Sec.3 ($\la=\La=0$) and of Sec.4($\la=0$).

\section{Evaluation of Coefficients and Numerical
Check of Analytic Results }
We present here the results of concrete values
of c's and d's for two {\it input} values $\vz=1.0$ and $\vz=1.6$.
We solve constraints (\ref{sol7}) by taking 
the first 7 terms (up to n=6th order). 
The most important point 
is to confirm the convergence of the infinite
series 
$\sum_{n=0}^\infty\frac{c_{2n+1}}{(2n+1)!}$, and 
$\sum_{n=0}^\infty\frac{d_{2n+1}}{(2n+1)!}$, 
which guarantees the present 
boundary condition. The 6th-order approximation calculation
gives the vacuum parameters as
\begin{eqnarray}
\vz=1.0(\mbox{input})\com\q \La=-0.272\com\q \la=2.88\com\nn
\vz=1.6(\mbox{input})\com\q \La=-1.50825\com\q \la=1.49925
\pr
\label{ana1}
\end{eqnarray}
\footnote{
The constraint is satisfied as
$
(1-\sum_{n=0}^6\frac{c_{2n+1}}{(2n+1)!})^2+
(1-\sum_{n=0}^6\frac{d_{2n+1}}{(2n+1)!})^2
\ <\ 1.1\times 10^{-7}\ \ (v_0=1.0),\ 1.6\times 10^{-9}\ \ (v_0=1.6)
$.
In the search of the solution $(\La,\la)$, the mesh-size of the parameter-space
determines the "precision". The high precision for the $\vz=1.6$
case shows the mesh-size is taken much smaller than
that of $\vz=1.0$.
}
The input value $\vz=1.6$ is quite near the case of \cite{SI0003}, 
and the above results of
$\La$ and $\la$ are consistent with the previous results.
Both cases give similar behaviours, hence we present
further results only for $\vz=1.0$. 
The obtained values of the coefficients are shown in Fig.1.
\begin{figure}
\epsfysize=6cm\epsfbox{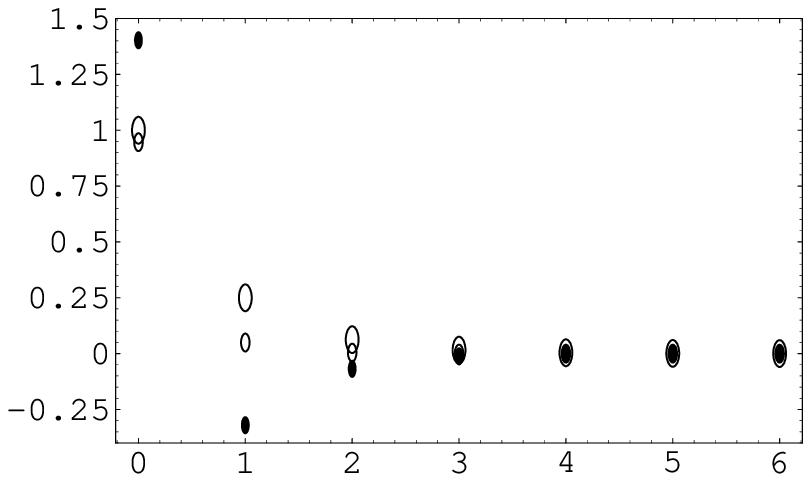}   
   \begin{center}
Fig.1\ 
The values of $c_{2n+1}/(2n+1)!$(blob),$d_{2n+1}/(2n+1)!$
(small circle).
The large circles show $(0.25)^n$. ( $ n=0,1,\cdots,6$ .) 
$\vz=1.0$(input).
   \end{center}
\end{figure}
In the figure we also plot the data from
the geometrical series: 
$1/(1-x)=1+x+\cdots$ at $x=0.25$.  
Comparing them,
we can recognize the (rapid) convergence
of the coefficient-series (up to this approximation order). 
Note that two series, 
$\{\frac{c_{2n+1}}{(2n+1)!}\}$ and 
$\{\frac{d_{2n+1}}{(2n+1)!}\}$, are 'oscillating'.
Using these results, the analytical results of
the scalar field $\Phi(y)$ and the warp factor $\si'(y)$, 
(\ref{sol5}), are shown in Fig.2.
\footnote{
We use the truncated version of the expressions (\ref{sol5})
by the first seven terms.
}
\begin{figure}
\epsfysize=6cm\epsfbox{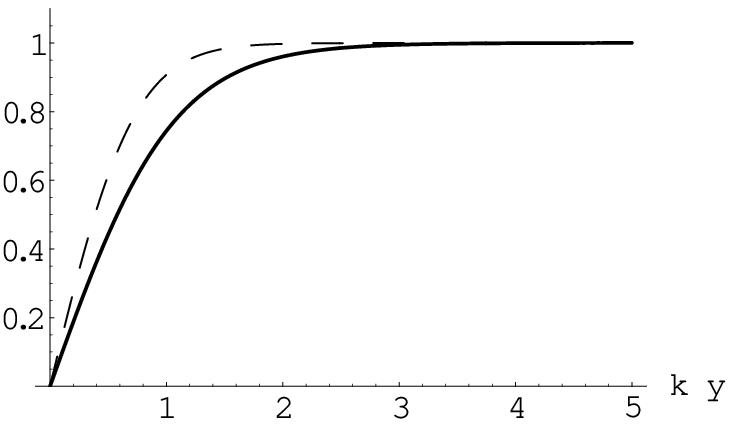}   
   \begin{center}
Fig.2\
The analytic result of $\si'/\om$(dashed line) and
$\Phi/\vz$(normal line). 
Both are odd with respect to $y\change -y$. 
The graphs are depicted by using (\ref{sol5}) with the 6-th order approximation. 
$\vz=1.0$(input).
The horizontal axis is $k y$. 
   \end{center}
\end{figure}
The (5D) Riemann scalar curvature is also shown in Fig.3. 
\begin{figure}
\epsfysize=6cm\epsfbox{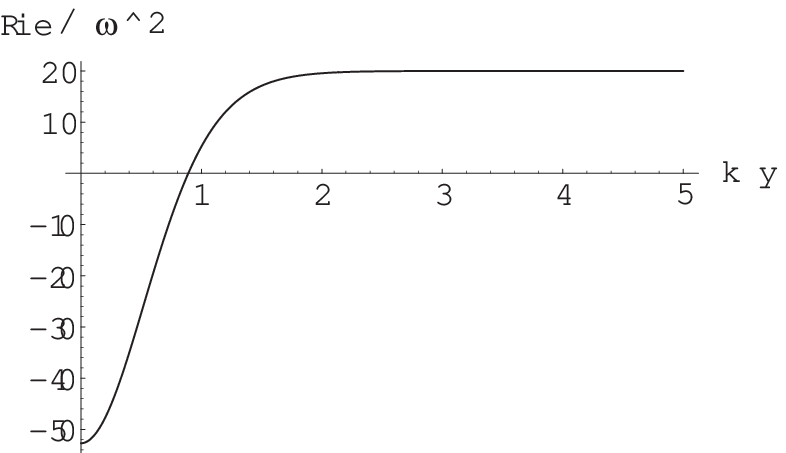}   
   \begin{center}
Fig.3\ 
(5D) Riemann scalar curvature $\Rhat/\om^2$
in the 6th order approximation. 
It even with respect to $y\change -y$.
The horizontal axis is $k y$. $\vz=1.0$(input).
   \end{center}
\end{figure}
It shows the curvature is {\it negative} inside
the wall, whereas {\it positive} outside. There
exist two points, $ky\approx \pm 1$, where
the curvature {\it vanishes}. 
We see the present solution
is {\it non-singular everywhere}. 
The presence of a dip around $y=0$ (Fig.3) clearly says
that the domain wall exists there. 

The kink b.c. for the Higgs field (\ref{sol3})
guarantees the stability of the present solution. It reflects
the b.c. for $\si'$ as specified by the parameter
value (\ref{sol4}). We can also see the stability from the
behaviour of $\Rhat$ (Fig.3) as follows. From the expression
(\ref{sol5}) and the b.c. (\ref{sol7}), $\Rhat$ has the 
following b.c. in the IR region.
\begin{eqnarray}
|\si'|\ra \om\com\q
\si''\ra 0\com\q
\frac{\Rhat}{\om^2}\ra 20\com\q
\mbox{as }\q |y|\ra \infty 
\pr
\label{ana2}
\end{eqnarray}
We can also see, from (\ref{sol5}), the b.c. in the UV region
is
\begin{eqnarray}
\si'\ra 0\com\q
\si''\ra \om k c_1\com\q
\frac{\Rhat}{\om^2}\ra -\frac{8k}{\om}c_1\com\q
\mbox{as }\q y\ra 0 
\com
\label{ana3}
\end{eqnarray}
where $c_1>0$ from the b.c. of $\si'$. Due to the continuity,
$\Rhat$ must have a dip around the origin with a finite thickness.
\footnote{
For the input $\vz=1.0$, the obtained values, 
$\La/k^2M^3=-0.272, c_1=1.4$, 
give $-8kc_1/\om=-52.6$ which is shown in Fig.3.
}

As the coupled differential equations for 
$\Phi(y)$ and $\si'(y)$, 
the equations (\ref{sol2a},\ref{sol2b}), 
have the standard form
of the numerical analysis, that is, Runge-Kutta method. 
We can numerically solve them {\it without any ansatz}
about the form of the solution.
In this numerical analysis, 
the following two points are important:\ 
1) the choice of three parameters $\vz, \la$ and $\La$;\ 
2) the choice of the initial conditions,
$\si'(y=0)$ and $\Phi(y=0)$. 
As for the point 1) we can borrow the values obtained 
in the 6th order approximation (\ref{ana1}).
As for 2), 
due to the required $Z_2$ symmetry
(the oddness under $y\ra -y$) for $\Phi(y)$ and $\si'(y)$ (\ref{sol5}),
we can take $y=0$ as the initial point of $y$ and 
the initial conditions $\Phi(0)=\si'(0)=0.0$. 
The numerical result is shown in Fig.4. 
\begin{figure}
\epsfysize=6.5605cm\epsfxsize=10.163cm\epsfbox{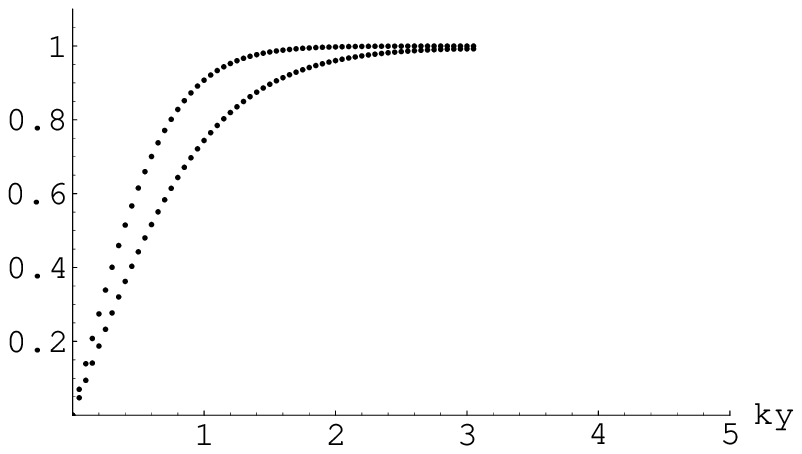}
   \begin{center}
Fig.4\
The numerical results for $\si'/\om$(up) and
$\Phi/\vz$(down). They are obtained by Runge-Kutta method.
One step value along $k y$-axis is 0.05. About 65 points are
plotted for each line in the figure. 
The initial point is $y=0$. 
The horizontal axis is $k y$. 
   \end{center}
\end{figure}
It shows the analytic solution of Fig.2, based on Sec.5, 
is reproduced very well. 
The numerical output data stop at $ky\sim 3.0$ 
with producing imaginary values.
This occurs because keeping the {\it positivity}, 
${\Phi'(y)}^2\geq 0$, in the numerical analysis
becomes so stringent
in the infrared region. The quantity becomes
so small in that region and vanishes at $k|y|=\infty$. 
We understand that further higher order
calculation is required
for the values of $\la$ and $\La$ (for an input value $\vz$)
in order to extend the valid region furthermore.

\section{Discussion and conclusion}
The assumption about the convergence of the series (\ref{sol5}), 
which was assumed in \cite{SI0003}, is strongly confirmed.
The present result of the 6th order calculation
does not so much 
deviate from that of the 2nd order one in \cite{SI0003}.
It says 
the truncation approximation of (\ref{sol7}) is valid.

We point out some results which are potentially
important in phenomenology. The cosmological term
has both the {\it upper} and the {\it lower} bound (\ref{sol10}).
It is expected to be useful when we fix the parameters
$\la, \vz$ and $\La$. The sign change of the curvature
near the "surface" $k|y|\approx 1$ (Fig.3) could become an
important check point of the confirmation of this model.

From the point of singularity resolution\cite{JPP9911},
the curvature singularity appearing in Sec.3 and Sec.4
is resolved in Sec.5 by a sort of "deformation"
(adding the potential terms appropriately). 
In the procedure two constraints appear among
the three vacuum parameters.

The present standpoint is that the domain configuration
should be realized in the non-singular geometry. The 
approach based on the singular geometry, such as
the original one\cite{RS9905}, can be regarded as a temporary
stage of the development. The singularity, often
expressed by the $\del$-function introduced by hand, 
is expected to be derived
by some definite limiting procedure, say, the thickness goes
to zero:\ $k^{-1} \ra 0$. 
If the string or the D-brane is the fundamental
constituents of nature, such extended objects are strongly
expected to behave {\it smoothly} in the UV-region. 
Because the domain world physics can be regarded as a transitive
approach from the field theory to the string-brane theory,
we believe seeking non-singular solutions is important
in the development.

\vspace{2cm}

\vs 1
\begin{flushleft}
{\bf Acknowledgment}
\end{flushleft}

The author thanks M. Abe, N. Ikeda and K. Nagao
for some comments and discussions at the YITP-workshop on
quantum field theory (Jul.16-19,2001, Kyoto).

\vs 1


\end{document}